\documentclass[aps,preprintnumbers,prd,twocolumn,superscriptaddress]{revtex4}

\usepackage{epsfig,latexsym,cancel,amssymb,amsmath}
\usepackage{graphicx}
\usepackage{epstopdf}
\usepackage{hyperref, graphicx, slashed, xcolor, amsmath}

\allowdisplaybreaks

\definecolor{red}{rgb}{1,0,0}

\newcommand{\beq}{\begin{equation}}
\newcommand{\eeq}{\end{equation}}
\newcommand{\bea}{\begin{eqnarray}}
\newcommand{\eea}{\end{eqnarray}}

\begin{document}

\title{ Microlensing effects of wormholes
	associated to blackhole spacetimes }
\author{Ke Gao}
\author{Lei-Hua Liu}
\email{liuleihua8899@hotmail.com}
\affiliation{Department of Physics, College of Physics,
	Mechanical and Electrical Engineering,
	Jishou University, Jishou 416000, China}
\author{Mian Zhu}
\email{mzhuan@connect.ust.hk}
\affiliation{Department of Physics, The Hong Kong University of Science and Technology, Clear Water
	Bay, Kowloon, Hong Kong, P.R. China}
\affiliation{Jockey Club Institute for Advanced Study, The Hong Kong University of Science and Technology, Clear Water Bay, Kowloon, Hong Kong, P.R. China}
\affiliation{Faculty of Physics, Astronomy and Applied Computer Science, Jagiellonian University, 30-348 Krakow, Poland}

\begin{abstract}

In this paper, we investigate the microlensing effects of wormholes
associated to black hole spacetimes. Specifically, we work on three typical wormholes (WH): Schwarzschild WH, Kerr WH, and RN WH, as well as their blackhole correspondences. We evaluate the deflection angle upon the second order under weak field approximation using Gauss-Bonnet theorem. Then, we study their magnification with numerics.We find that a Kerr WH could lead to multi peaks in the magnification with certain parameters in the prograde case, while a Kerr BH predicts one peak. Therefore, the multi-peak feature of can be used to distinguish the Kerr WH from other compact objects. We also find that the magnification of RN BH will be one peak compared to RN WH, in which the magnification of RN WH is negative in some situations. For other cases, the behavior of magnification from wormholes and their corresponding blackholes is similar. Our result may shed new light on exploring compact objects through the microlensing effect.

\end{abstract}

\maketitle

\bigskip

\section{Introduction}
\label{introduction}
Wormhole (WH) \cite{Morris:1988cz,Einstein:1935tc,Fuller:1962zza,Bronnikov:1973fh, Ellis:1973yv} is a hypothetic geometric structure connecting two otherwise remote regions. Wormholes may permit faster-than-light travel and time travel \cite{Morris:1988tu}. Furthermore, in the framework of General Relativity, the construction of traversable wormholes requires the violation of Null Energy Condition (NEC) \cite{Hochberg:1998ii, Hochberg:1998ha}, and exotic matters beyond our current scope are necessary. Hence, the existence of wormholes may improve our understanding of new physics and it is important to study wormhole physics. 

Gravitational lensing is a promising approach to search for wormholes \cite{Narayan:1996ba,Bartelmann:1999yn, Perlick:2004tq}. In the literature, the lensing effect of a wormhole is extensively studied \cite{Safonova:2002si,TejeiroS:2005ltc,Nandi:2006ds,Abe:2010ap,Toki:2011zu,Yoo:2013cia,Takahashi:2013jqa,Izumi:2013tya,Kuhfittig:2013hva,Nakajima:2014nba,Tsukamoto:2016zdu,Shaikh:2017zfl,Asada:2017vxl,Shaikh:2018oul,Shaikh:2019itn,Javed:2019qyg,Shaikh:2019jfr,Dai:2019mse,Simonetti:2020ivl,Bambi:2021qfo,Godani:2021aub,Liu:2022lfb,Qiao:2022nic,Petters:2010an}. In the weak field region, the wormholes can be treated as a dark compact object \cite{Cardoso:2019rvt}, and a large variety of the wormholes mimics the black holes (BHs) \cite{Damour:2007ap,Tsukamoto:2012xs,Abdikamalov:2019ztb}.Up to now, it would be difficult to distinguish them by astrophysical observations \cite{Berti:2015itd,Barack:2018yly}. Thus, people are motivated to distinguish wormholes from other compact objects with various techniques, see e.g.  \cite{Amir:2018pcu,Kasuya:2021cpk,Vagnozzi:2022moj,Shaikh:2019hbm,Karimov:2020fuj,Jusufi:2018gnz,Konoplya:2016hmd}.

In this paper, we proceed with a slightly different approach. The lensing effect of simple objects is well-studied. For example, the lensing physics of massless Ellis-Bronnikov wormhole (EBWH) can be done even in the strong field region \cite{Abe:2010ap,Tsukamoto:2016qro,Tsukamoto:2016jzh}, while the lensing effect of more generic metric might be hard to proceed. Hence, we work in the weak field approximation (that is, the impact parameter $b_I$ is much larger than the intrinsic parameters of the wormhole/blackhole) for simplicity. Then, we study the lensing effect beyond the leading order, and see if the higher order contribution may help to distinguish them. For convenience, we adopt the technique introduced in \cite{Gibbons:2008rj,Gibbons:2008zi,Werner:2012rc}, where the deflection angle is evaluated through the Gaussian-Bonnet theorem (GBT). The GBT formalism is wildly applied to study the deflection angle of various wormhole models since it elaborate the physics in a geometric way \cite{Jusufi:2017vta,Jusufi:2017mav,Jusufi:2017drg,Ovgun:2018fnk,Ono:2018ybw,Jusufi:2018kmk,Ovgun:2018prw,Ovgun:2020yuv}.

We organize the paper as follows. In section \ref{sec:formalism}, we briefly introduce wormhole physics, gravitational lensing physics, and how to evaluate the deflection angle using GBT formalism. We explicitly show how GBT formalism works by using the massless EBWH as an example in section \ref{sec:ellis}. We then study the magnification of Schwarzschild case in section \ref{sec:Sch}, Kerr case in section \ref{sec:Kerr}, and RN case in section \ref{sec:RN}. We find that it is possible to distinguish Kerr WH and BH through the difference of magnitudes between their gentle peaks and main peaks. We conclude in section \ref{sec:conclusion}.

\section{Basic formalism}
\label{sec:formalism}
In this section, we firstly review the basics of wormhole physics. After that, we discuss the gravitational lensing physics, and show how to use the GBT formalism to study the lensing physics.

\subsection{Wormhole physics}
\begin{figure}[h]
	\centering
	\includegraphics[scale=0.6]{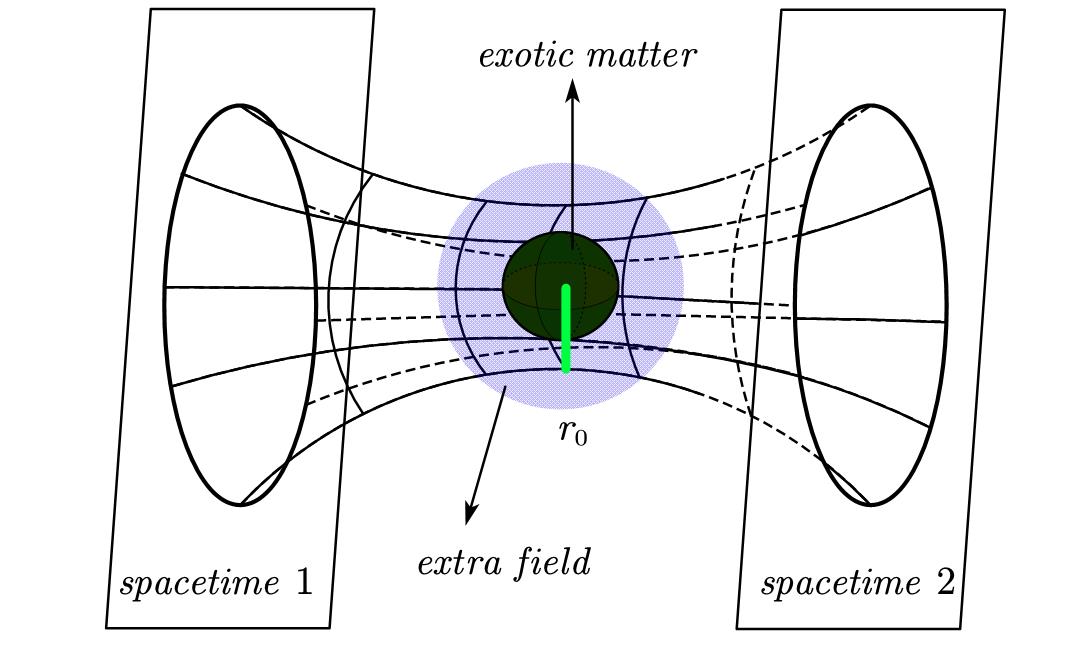}
	\caption{A wormhole connects two spacetime, the shadow part represents the structure of throat and $r_0$ is the radius of throat. }
	\label{throat structure}
\end{figure}

For simplicity, we shall consider static spherically symmetric wormholes only. We start with the Morris-Throne wormhole \cite{Morris:1988cz,Morris:1988tu}. The metric is given by
\begin{equation}
ds^2=-e^{2\Lambda \left( r \right)}dt^2+\frac{dr^2}{1-b\left( r \right) /r}+r^2 d\Omega_2^2,
\label{eq:MSWH}
\end{equation}
where $d\Omega_2^2$ is the metric of a unit 2-sphere. The function $\Lambda\left( r \right)$ and $b\left( r \right)$ are the redshift function and shape function, respectively. The wormhole structure is characterized by its throat that connects two regions of spacetime. We illustrate a typical wormhole structure in figure \ref{throat structure}. In the weak field limit, the lensing always happens at one side.
The wormhole throat is defined such that the geometry is the minimality in the embedded spacetime. To understand the flare-out condition, it is better to embed into a lower geometry that means $t=\rm constant$ and set $\theta=\pi/2$ of solid angle $\Omega_2$, then one can easily obtain $ds^2_2=\bigg(1-b(r)/r\bigg)^{-1}dr^2+r^2d\phi^2=dz^2+dr^2+r^2d\phi^2$. Thus, one can find $dr/dz=\pm \frac{b}{r-b}$. Finally, one can easily derive the flare-out condition by implementing $\frac{d}{dz}\frac{dr}{dz}=\frac{dr}{dz}\partial_r (\pm \frac{b}{r-b})$ as follows,
\begin{equation}
b(r_0) = r_0 ~,~ \frac{b(r) - r b'(r)}{2b(r)^2} > 0 ~,
\end{equation}
with $b'(r) \equiv db(r)/dr$ and $r=r_0$ labels the location of the throat. We see that the structure of the wormhole is solely determined by the shape function $b(r)$. We also impose the asymptotic flatness, which sets $\displaystyle\lim_{r \to \infty} b(r)/r = 0$. Finally, a traversable wormhole should have no horizon, i.e. $g_{tt} \neq 0$ everywhere. In the metric \eqref{eq:MSWH} this translates into $\Lambda (r)$ being finite everywhere.

\subsection{Gravitational lensing}
\begin{figure}[t]
	\centering
	\includegraphics[scale=0.40]{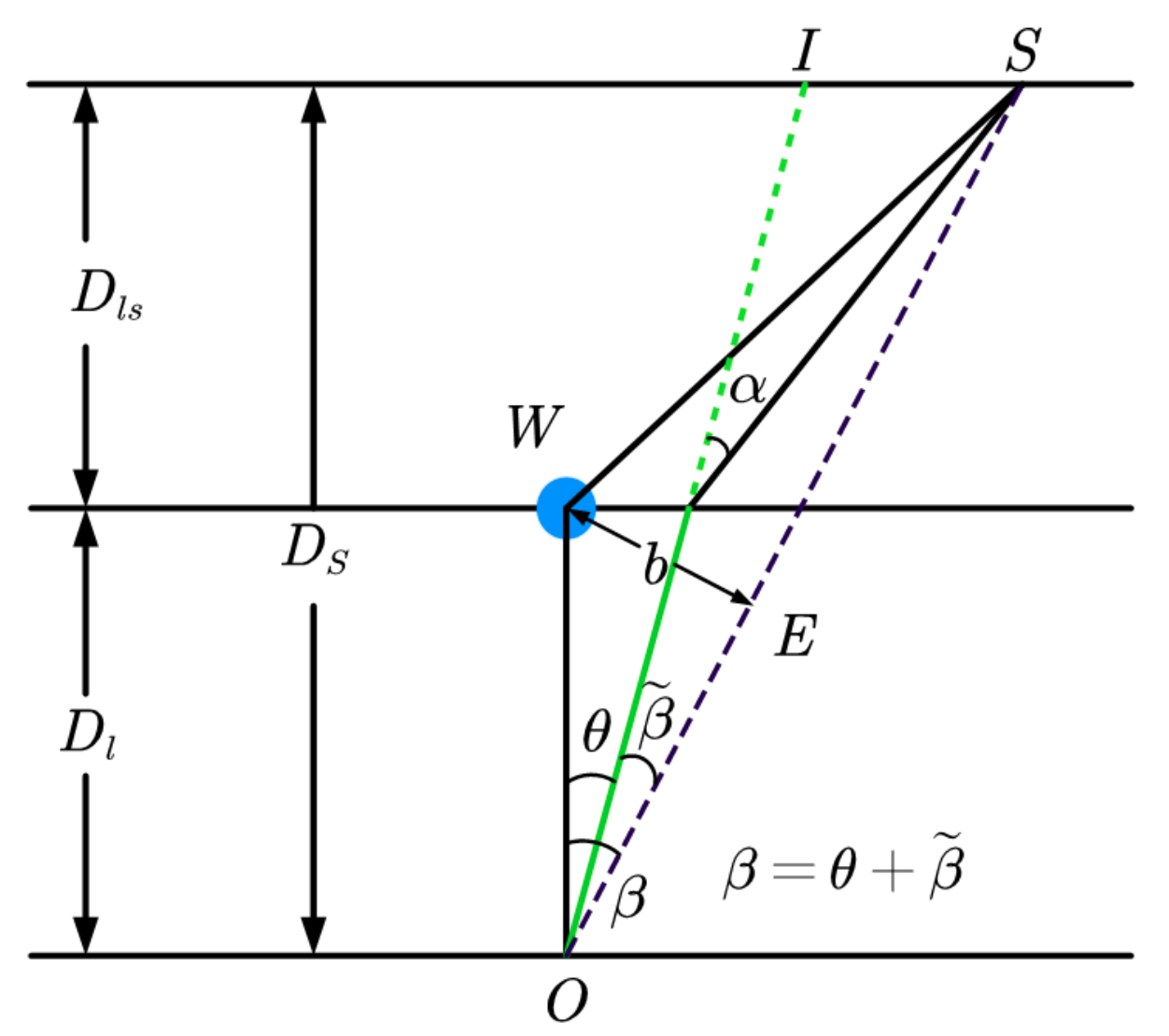}
	\caption{The lensing geometry. Here $W$ corresponds to the WH as a lens. $O$ and $S$ are observers and sources, respectively. $I$ is the image of $S$, and $\theta$ is the angle of the corresponding image and the wormhole. $\alpha$ is the deflection angle, and $\beta$ is the angle between the wormhole and the light source. $D_{l}, D_{ls}$ and $D_{s}$ are angular diameter distances. Other quantities are auxiliary.}
	\label{fig:my_labal 2}
\end{figure}

A typical gravitational lensing geometry is illustrated in figure \ref{fig:my_labal 2}. For convenience, let's focus on an infinitesimal source. The images observed will be magnified or demagnified due to the change of cross-section of a bundle of rays. The magnification is determined by the ratio between the solid angles
\begin{equation}
\label{eq:magnification}
| \mu | = \frac{d\omega_i}{d\omega_s} = \left| \frac{\beta}{\theta} \frac{d\beta}{d\theta} \right|^{-1} ~.
\end{equation}

The lensing geometry in figure \ref{fig:my_labal 2} gives the lens equation
\begin{equation}
\label{eq:lenseq}
\beta =\theta -\frac{D_{ls}}{D_s}\alpha.
\end{equation}
Hence, if we work out the deflection angle $\alpha$ as a function of $\theta$, we can use \eqref{eq:lenseq} to get $\beta(\theta)$ and finally the magnification $|\mu|$ from \eqref{eq:magnification}, which is an important observable in astrophysics.

Finally, the lens equation \eqref{eq:lenseq} may admit more than one solution of $\beta(\theta)$, corresponding to multiple images. For simplicity, we shall consider the microlensing case, where the separation of images is too small to be resolved by existing telescopes. In this case, we observe the combined light intensity, i.e. the observed magnification should be the summation of magnifications of each image:
\begin{equation}
|\mu_{total}| = \sum_i |\mu_i| ~.
\label{eq:total magnification}
\end{equation}

\subsection{Formalism with Gauss-Bonnet Theorem}

 The deflection angle can be evaluated by GBT, which is given by 
\begin{equation}
\label{eq:alphaGBT}
\alpha =-\int{\int_{D_2}{K dS}},
\end{equation}	
where $D_2$ is a domain outside the light ray, the $K$ stands for the Gaussian optical curvature and $dS$ is the elementary surface area of the optical geometry.
\begin{figure}[h]
	\centering
	\includegraphics[scale=0.804]{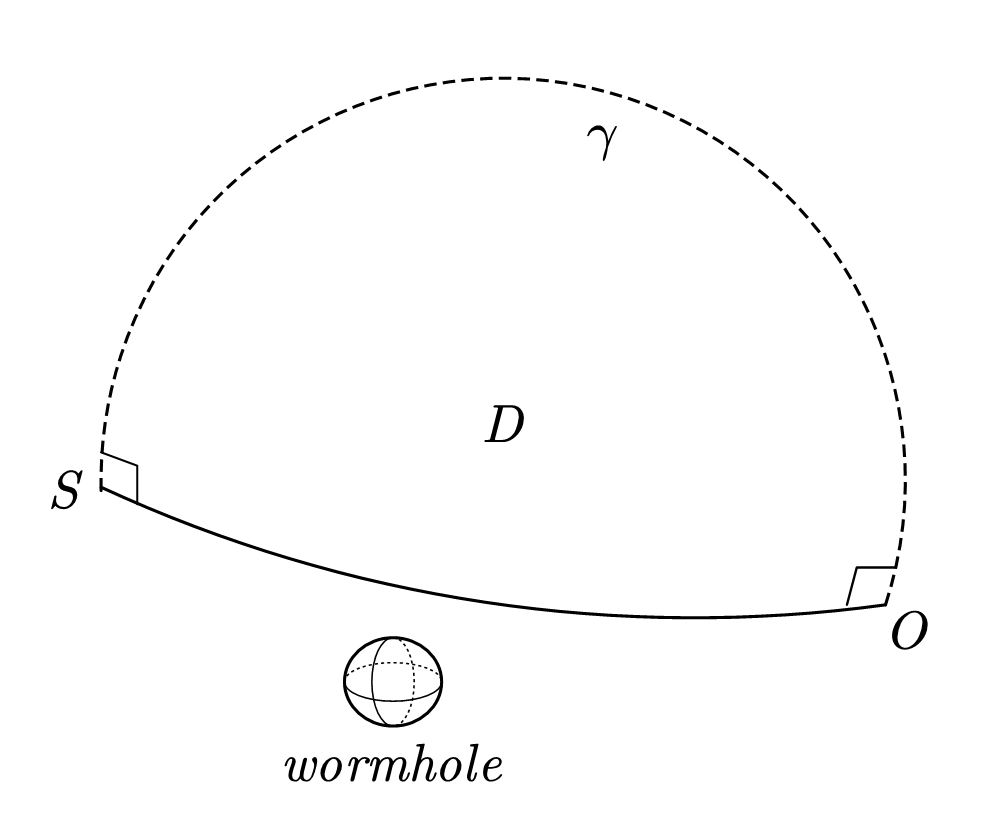}
	\caption{The geometry of lensing. $OS$ is the geodesic line and $D$ is a domain for the integration, $\gamma$ is a line which is vertical point $S$ and $O$.}
	\label{fig:GBTgeo}
\end{figure}

The formula \eqref{eq:alphaGBT} comes from the Gauss-Bonnet theorem, which in the lensing geometry \ref{fig:GBTgeo} becomes
\begin{equation}
\iint\limits_D{K}\,\,dS +\oint\limits_{\partial D}{\kappa \,\,dt+\sum_i{\alpha _i}}= 2\pi \chi(D) ~,
\label{gaussian bonnet}
\end{equation}
where the domain $D$ is simply-connected and $\partial D$ is its boundary, $\sum_i{\alpha_i}$ is the sum of the exterior angle when taking the domain as a polygon and $\chi(D)$ is the Euler characteristic number, which $\chi(D)=1$ in our case . The function $\kappa$ is the curvature of the geodesics. Since the line $SD$ is the geodesic line, its corresponding integration of $\kappa$ along $SD$ is vanishing, then 
\begin{equation}
	\int\int_{D_2}KdS+\int_{\gamma}\kappa dt+\sum_i\alpha_i=2\pi.
\end{equation}
where $D_2$ is the domain except the line $SD$. As Fig. \ref{fig:GBTgeo} shows, line $\gamma$ is vertical $SD$, which means that 
\begin{equation}
\sum_i\alpha_i=\frac{\pi}{2}(S)+\frac{\pi}{2}(O)=\pi,
\end{equation}
the sum of exterior angles changes into the sum of two vertical angles. Afterward, 
one can do the integral transformation as follows,
\begin{equation}
\kappa dt=\kappa\frac{dt}{d\phi}d\phi.
\end{equation}
For the line $\gamma$, one can set $\kappa\frac{dt}{d\phi}=1$, then we have 
\begin{equation}
\int\int_{D_2}KdS+\int_{\phi_O}^{\phi_S}d\phi+\pi=2\pi.
\end{equation}
For the very large scale, the geodesic line $OS$ can be approximated to a straight line, the angle can be spanned as $\pi+\alpha$ ($\alpha$ is the deflection angle),
\begin{equation}
\int\int_{D_2}KdS+\int_{0}^{\pi+\alpha}d\phi+\pi=\int\int_{D_2}KdS+\pi+\alpha+\pi=2\pi.
\end{equation}
Finally, one can get \begin{equation}
\alpha=-\int\int_{D_2}KdS.
\end{equation}

\section{Massless EBWH as an example}
\label{sec:ellis}
In this section, we illustrate how the GBT formalism \eqref{eq:alphaGBT} works for the massless EBWH \cite{Ellis:1973yv,Bronnikov:1973fh}. The metric is
\begin{equation}
ds^2=-dt^2+dr^2+\left( r^2+r_0^2 \right) d\Omega_2^2, 
\label{eq:Ellis}
\end{equation}	
and after a coordinate transformation $\rho = \sqrt{r^2 + r_0^2}$, the metric returns to a Morris-Throne type:	
\begin{equation}
ds^2 = -dt^2 + \frac{d\rho^2}{1 - r_0^2/\rho^2} + \rho^2 d\Omega_2^2 ~, 
\end{equation}	
and it's easy to see that the throat radius is $r_0$.	

For a photon, the geodesic equation is $ds^2 = 0$. Also, we may simplify the problem by working in the equatorial plane with $\theta = \pi/2$. The geodesics of a photon is then described by
\begin{equation}
\label{eq:geoEllis}
dt^2 = dr^2 + (r^2 + r_0^2) d\varphi^2 ~.
\end{equation}

Now we define auxiliary functions $du = dr$, $\zeta(u) = \sqrt{r^2 + r_0^2}$, such that equation \eqref{eq:geoEllis} becomes
\begin{equation}
dt^2=h_{ab}d\lambda ^ad\lambda ^b=du^2+\zeta ^2\left( u \right) d\varphi ^2.
\label{eq:hab}
\end{equation}
The Gaussian optical curvature is then
\begin{equation}
\mathcal{K} =-\frac{1}{\zeta \left( u \right)}\left[ \frac{dr}{du}\frac{d}{dr}\left( \frac{dr}{du} \right) \frac{d\zeta}{dr}+\left( \frac{dr}{du} \right) ^2\frac{d^2\zeta}{dr^2} \right] .
\label{eq:Gausscurvature}
\end{equation}
Using the formulae \eqref{eq:alphaGBT}, the deflection angle is
\begin{equation}
\alpha =-\int_0^{\pi}{\int^{\infty}_{b/\sin\varphi}{\mathcal{K} \sqrt{\det h_{ab}}}}drd\varphi.
\label{eq:alphaEllisGB}
\end{equation}
Here, the $r$ integral ranges from the source to the observation. Using the lens geometry in figure \ref{fig:my_labal 2} and with the help of equations \eqref{eq:hab} and \eqref{eq:Gausscurvature}, we finally get

\begin{align}
\alpha = \pi - 2 K \left( \frac{r_0^2}{b^2} \right) ~,
\label{deflection ang of ellis}
\end{align}

where $K(k)$ is the complete elliptic function of the first kind.
In weak field approximation, the deflection angle simplifies to
\begin{equation}
\alpha = \frac{\pi}{4} \left( \frac{r_0}{b_I} \right)^2 - \frac{9\pi}{64} \left( \frac{r_0}{b_I} \right)^4 + \mathcal{O} \left( \frac{r_0}{b_I} \right)^6 ~,
\end{equation}
in agreement with \cite{Nakajima:2012pu,Jusufi:2017gyu}.  Unfortunately, the ADM mass of \eqref{eq:Ellis} is zero, thus one cannot find its corresponding blackhole.

\section{Schwarzschild Case}
\label{sec:Sch}

The metric of Schwarzschild wormhole is \cite{Damour:2007ap}, 
\begin{equation}
ds^2=-\left( 1-\frac{2M}{r}+\lambda ^2 \right) dt^2+\frac{dr^2}{1-\frac{2M}{r}}+r^2d\varOmega ^2,
\label{sch wormhole}
\end{equation}
where $\lambda$ is a parameter and M is the mass. The Schwarzschild BH is restored when $\lambda=0$. The throat is located at $r = 2M$. The Gaussian curvature \eqref{eq:Gausscurvature} of the metric \eqref{sch wormhole} is
\begin{equation}
	\mathcal{K}=\frac{6(\lambda^2+1)M^3-7(\lambda^2+1)r M^2+r^2(\lambda^2+2)M}{(-r+2M)r^4},
	\label{eq:Gaussiancur of schwar}
\end{equation}
Notice that this Gaussian curvature is an exact formula without using weak field approximation. The deflection angle up to the second order of $\frac{M}{b_I}$ is
\begin{equation}
	\alpha = \left( 4+2\lambda^2 \right) \frac{M}{b_I} +  \frac{7\pi}{4} (\lambda^2 + 1) \frac{M^2}{b_I^2},
	\label{eq:deflection angle of schwormhole}
\end{equation}
and the first term is consistent with Ref. \cite{Ovgun:2018fnk}. The deflection angle for Schwarzschild BH is recovered when $\lambda=0$.

\begin{figure}[ht]
	\centering
	\includegraphics[width=0.9\linewidth]{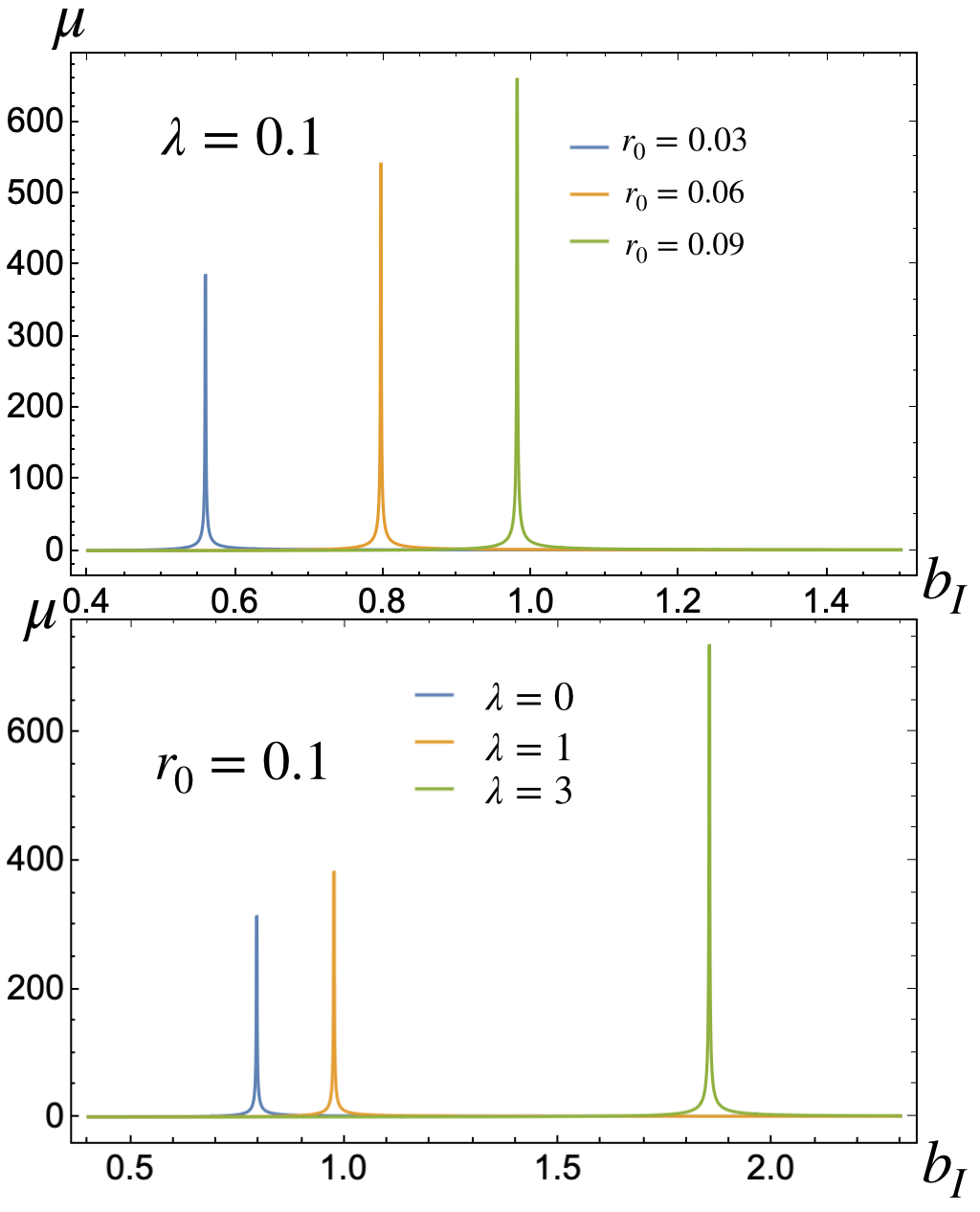}
	\caption{The upper panel shows the magnification with a fixed $\lambda=0.1$ and varying $r_0=0.03,~0.06,~0.09~\rm kpc$. The lower panel shows the magnification for various $\lambda$ with fixed $r_0=0.06~\rm kpc$. We have set $D_l=10~\rm kpc$ for both plots. The blue line in the lower channel corresponds to the Schwarzschild blackhole with $\lambda = 0$. }
	\label{fig:mag of schwarz1}
\end{figure}

We come to the magnification by numerically simulate the magnification as a function $\mu\equiv  \mu(r_0,\lambda,b_I)$. We use $r_0 = 2M$ instead of M as free parameters in figure \ref{fig:mag of schwarz1} for convenience. We set $D_l=10~\rm kpc$, which is a typical galaxy scale. Meanwhile, we also set $\frac{D_{ls}}{D_s}=\frac{1}{2}$ for simplicity. 

From figure \ref{fig:mag of schwarz1}, we see the peak of magnification will be enhanced as we increase the parameter $r_0$ and $\lambda$. A Schwarzschild BH ($\lambda = 0$) would have a minimal peak of magnitude compared to a Schiwarzschild WH ($\lambda > 0$). However, it's hard for use to make use of this fact, due to the difficulty to determine the mass of the lens object.

\section{Kerr CASE}
\label{sec:Kerr}

The Kerr wormhole is described by \cite{Bueno:2017hyj}
\begin{align}
	ds^2 & \nonumber = -\left(1-\frac{2Mr}{\Sigma}\right)dt^2-\frac{4M\vec{a}r\sin^2(\theta)}{\Sigma}dtd\phi+\frac{\Sigma}{\hat{\Delta}}dr^2 \\
	& +\Sigma d\theta^2+\left(r^2+a^2+\frac{2M\vec{a}^2r\sin^2\theta}{\Sigma}\right)\sin^2\theta d\phi^2 ~,
	\label{eq:kerr metric}
\end{align}
with 
\begin{equation}
\begin{split}
\Sigma=r^2+\vec{a}^2\cos^2(\theta) ~,~\hat{\Delta}=r^2-2M(1+\lambda^2)r+\vec{a}^2 ,
\label{eq:kerr other}
\end{split}
\end{equation}
where $\vec{a}\equiv \frac{J}{M}$ and $J$ is the angular momentum. The Kerr BH is recovered with $\lambda = 0$. One can always project $\vec{a}$ onto the equatorial plane as $a=\vec{a}\cos(\psi)$ where $\psi$ is the angle between the  equatorial plane and $\vec{a}$. After this transformation, all of calculation will be evaluated on the the equatorial plane that denotes $\theta=\pi/2$, in which we will keep $a$ for the later investigation. The radius of throat locates at $\Delta = 0$, or more explicitly
\begin{equation}
	r_+=M(1+\lambda^2)+\sqrt{M^2(1+\lambda)^2-a^2} ~.
\end{equation}	
In the weak field limit, we shall always work in the region $b_I \gg r_+$.

The deflection angle at the first order is already evaluated in Ref. \cite{Ovgun:2018fnk}:
\begin{equation}
	\alpha\approx \frac{2M(\lambda^2+2)}{b_I}\pm \frac{4Ma}{b_I^2} ~,
\end{equation}	
and the $\pm$ sign corresponds to the retrograde and the prograde light ray, respectively.

The infinitisimal area $d\sigma=\sqrt{\det h_{ab}}drd\varphi$ is
\begin{equation}
\small	 d\sigma=\sqrt{\frac{\Sigma^2}{\hat{\Delta}(\Sigma-2mr)}\bigg(r^2+a^2+\frac{2a^2mr}{\Sigma-2mr}\bigg)\frac{\Sigma}{(\Sigma-2mr)}}drd\varphi
	 \label{eq:area of kerr}
\end{equation}
Since Kerr WH rotates, we need to be more careful about $\kappa$, which becomes
\begin{equation}
	\kappa\approx \pm \left[-\frac{2aM}{r^3}+\frac{2M^2a\lambda^2}{r^4}-\frac{2aM^2}{r^4}+\mathcal{O}\left(\frac{1}{r^5}\right)\right] ,
	\label{eq:kerr geodesic cur}
\end{equation}
and the leading term is consistent with \cite{Ono:2017pie}. Here again the $\pm$ sign corresponds to the retrograde and the prograde case.
The Gaussian curvature is \cite{Ovgun:2018fnk} 
\begin{equation}
	\mathcal{K}\approx \frac{(\lambda^2+2)M}{r^3} ~.
	\label{eq:gaussian curvature of kerr}
\end{equation}
Now the expression for the deflection angle is
\begin{equation}
	\alpha=-\int_{\phi_S}^{\phi_R}\int_r^\infty K\sqrt{\gamma}drd\phi+\int_{l_S}^{l_R}\kappa dl,\
	\label{eq:deflection angle of kerr}
\end{equation}
where we set $l_R$ and $l_S$ to be infinity.  Finally we get
\begin{equation}
\small	\alpha_{\rm pro}=\frac{2M(\lambda^2+2)}{b_I}-\frac{4Ma}{b_I^2}+\frac{3M^2\pi(\lambda^2+2)}{2b_I^2}-\frac{aM^2\pi(\lambda^2+5)}{b_I^3}.
	\label{eq:prograde case of kerr}
\end{equation}
for the prograde case and
\begin{equation}
\small \alpha_{\rm re}=\frac{2M(\lambda^2+2)}{b_I}+\frac{4Ma}{b_I^2}+\frac{3M^2\pi(\lambda^2+2)}{2b_I^2}-\frac{3aM^2\pi(\lambda^2+1)}{b_I^3}.
\label{eq:retrograde case of kerr}
\end{equation}
for the retrograde case, in which the first order of deflection for the the prograde case and the retrograde case is consistent with \cite{Ono:2017pie,Ovgun:2018fnk}. Following their method, we perform our calculation on the equatorial plane. In \cite{Cai:2023ite}, they investigated the error of higher order to lensing effects of Ellis-Bronnikov wormhole, in which one can use their methods into our cases.

\begin{figure}[h]
	\centering
	\includegraphics[width=0.9\linewidth]{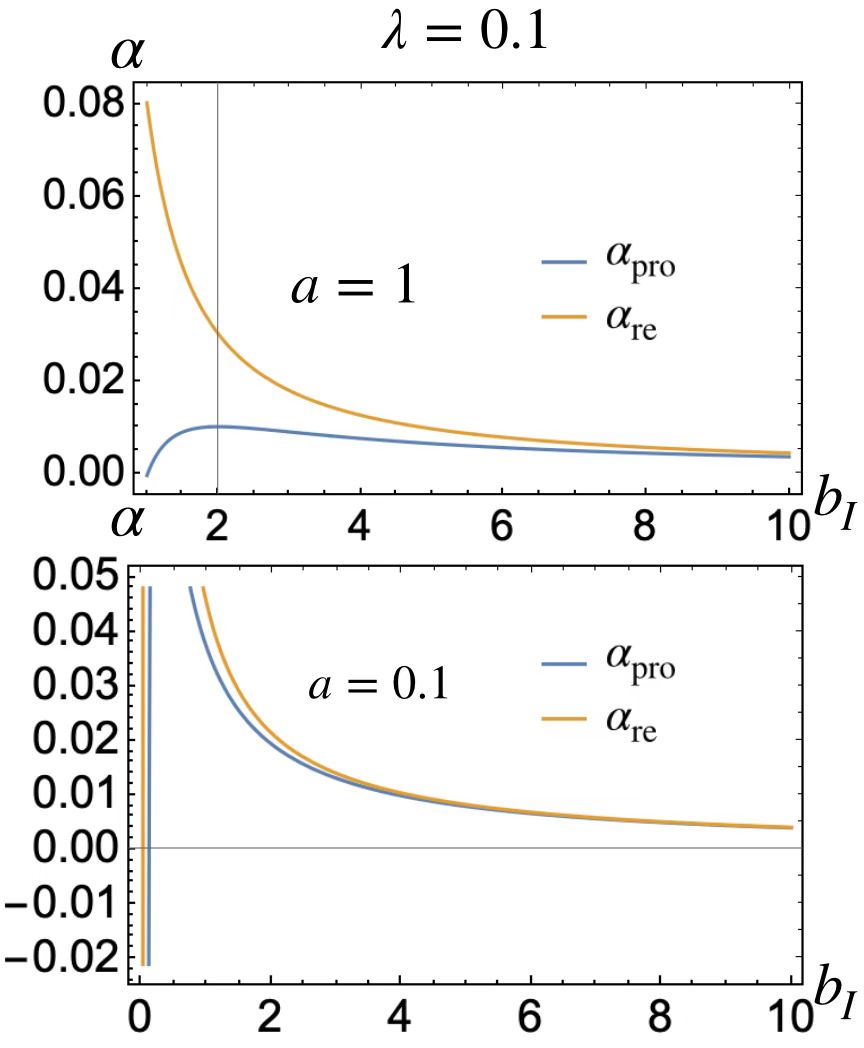}
	\caption{The deflection angle for prograde case (blue line) and retrograde case (orange line), respectively. For both cases, We have set $M=0.01$. The upper and lower channels correspond to $a=1$ and $a=0.1$, respectively.}
	\label{fig:deflection angle of kerr}
\end{figure}

We evaluate the deflection angle for each cases in figure \ref{fig:deflection angle of kerr}. We see that the prograde and retrograde cases differs significantly for large $a$, and almost indistinguishable for smaller $a$. Thus, it is important to work on the magnification with different value of $a$.

\begin{figure}[h]
	\centering
	\includegraphics[width=0.9\linewidth]{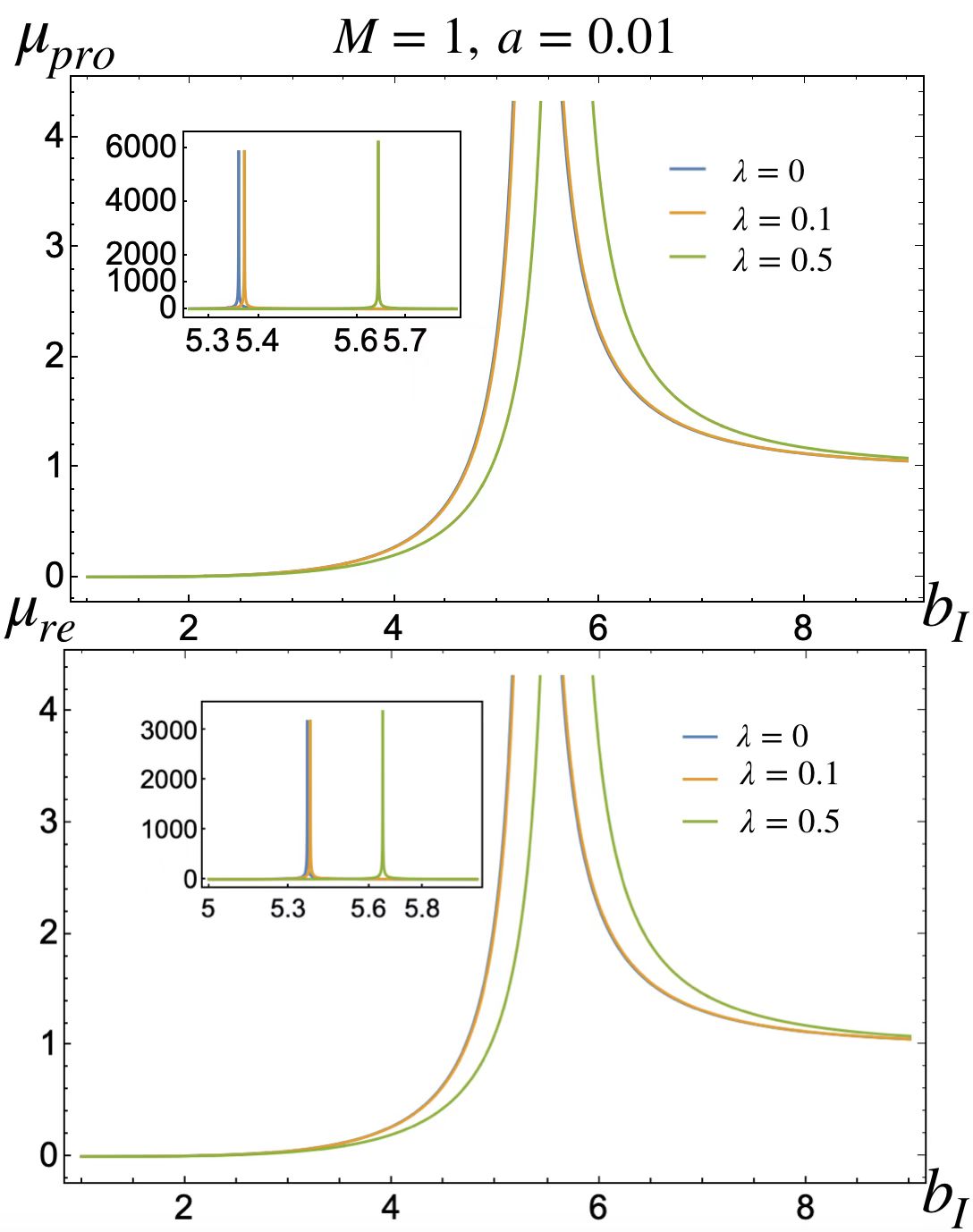}
	\caption{The magnification of Kerr metric  \eqref{eq:kerr metric} with $M \gg a$. We include the prograde case (upper channel) and retrograde case (lower channel). All of these units have unified and the value of $1.0~\rm kpc<b_I <9~\rm kpc$ ensures the weak field approximation. The maximal values of the magnification are around $6000$ and $3000$, respectively. }
	\label{fig:mag of kerr as m larger than a}
\end{figure}

We firstly come to the case $M \gg a$, since in this case the wormhole is very close to the Kerr black hole ($a=0$).
Figure \ref{fig:mag of kerr as m larger than a} shows the magnification in this case, and the magnification is almost the same for the prograde and retrograde case, as we expected from figure \ref{fig:deflection angle of kerr}.

The maximal order of this magnification is around $10^3$. The Kerr black hole ($a=0$) corresponds to the blue solid line in figure \ref{fig:mag of kerr as m larger than a}. Our numerical results show that the peak will appear with larger $b_{I}$ when we increase $\lambda$. However, the magnitude and the location of the peaks only depend on $\lambda$ slighty, so it would be hard to astrophysically distinguish them at current stage. Besides, the magnitude of the peaks in the prograde case is approximately twice of that in the retrograde case, which may potentially help us to distinguish them.

\begin{figure}[h]
	\centering
	\includegraphics[width=0.9\linewidth]{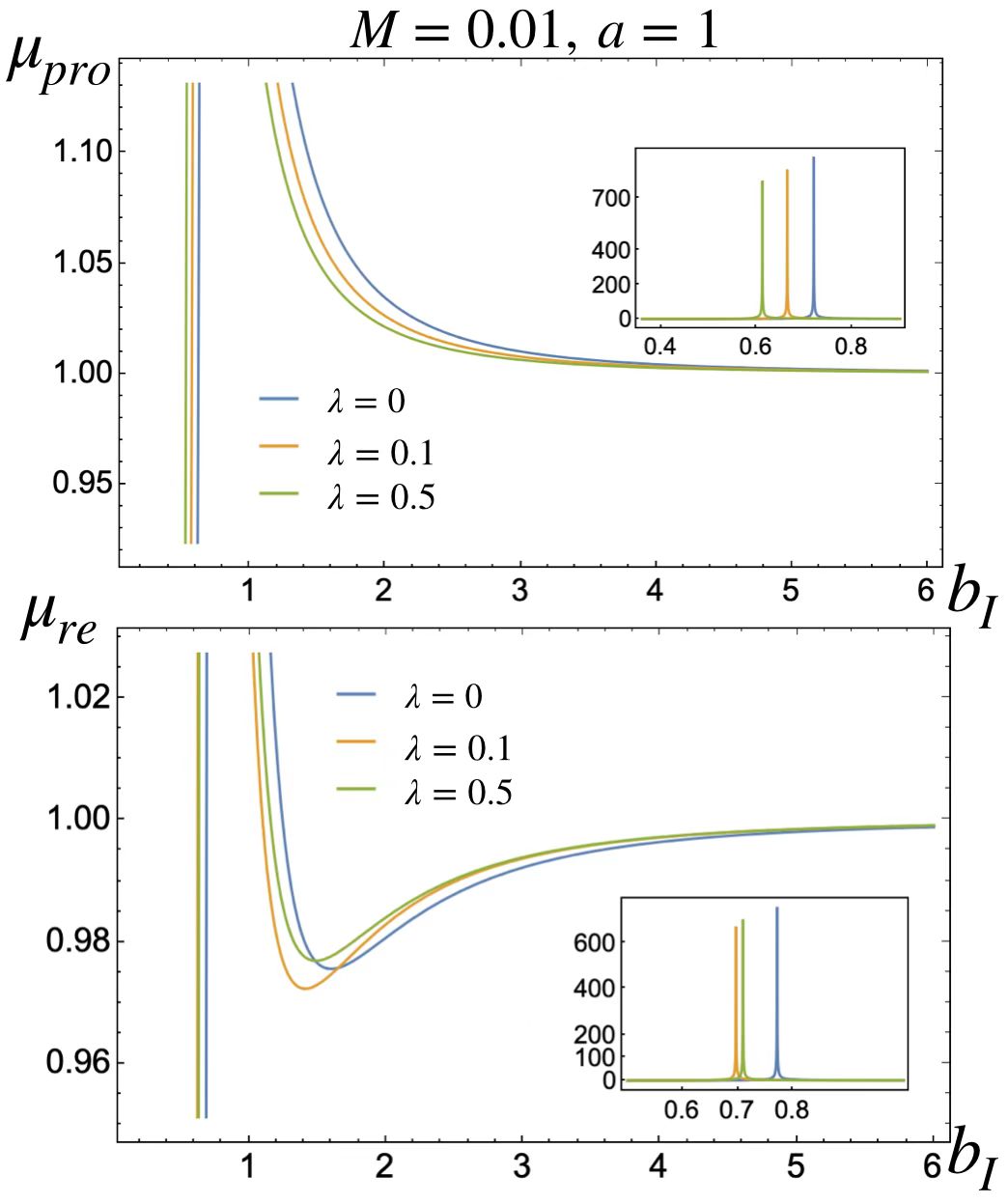}
	\caption{The magnification of Kerr metric  \eqref{eq:kerr metric} with $M \ll a$. We include the prograde case (upper channel) and retrograde case (lower channel). All of these units have been unified and the value of $0.1~\rm kpc<b_I <6~\rm kpc$ ensures the weak field approximation. The maximal order of the magnification is around $10^2$. }
	\label{fig:mag of kerr as a larger than m}
\end{figure}

Similar conclusion can be obtained in the $M \ll a$ case. Unfortunately, there is a tricky issue to address, since a Kerr BH is not stable when $M\ll a$. Thus, the gravitational lensing effect for such a Kerr BH may not be significantly relevant to observations. This situation is changed dramatically as Ref. \cite{Mazza:2021rgq} proposed a novel family of rotating blakholes, in which one can get their new metric by introducing $\tilde{r}^2=r^2+l^2$ ($l$ denotes the regularisation of the central singularity of blackhole.). In the metric building of \cite{Mazza:2021rgq}, the metric will be recovered into the Kerr metric as $l=0$. The most essential result is included in Fig. 1 of \cite{Mazza:2021rgq}, in which it shows that the Kerr BH will become the traversable Kerr WH as $a/M>1$ and Eq. (2.16) (l=0) of \cite{Mazza:2021rgq} is equivallent to our metric  \eqref{eq:kerr metric} with $\lambda=0$. Consequently, the blue line of Fig. \ref{fig:mag of kerr as a larger than m} shows the magnification of  traversable Kerr wormhole not the Kerr BH anymore. The other colorful lines is still the Kerr WH. There is an additional valley in the retrograde case, potentially enabling us to distinguish it from the prograde case. Moreover, the magnitude of peaks in each case is ten times smaller than that in the $M \gg a$ case, so we may have a change to learn the parameter $m/a$ by the amplitude of magnification. Unfortunately, our purpose is to distinguish WHs $\lambda > 0$ with their corresponding BHs $\lambda = 0$, and this is hard to achieved since the magnification varies slowly with $\lambda$.

\begin{figure}[h]
	\centering
	\includegraphics[width=0.9\linewidth]{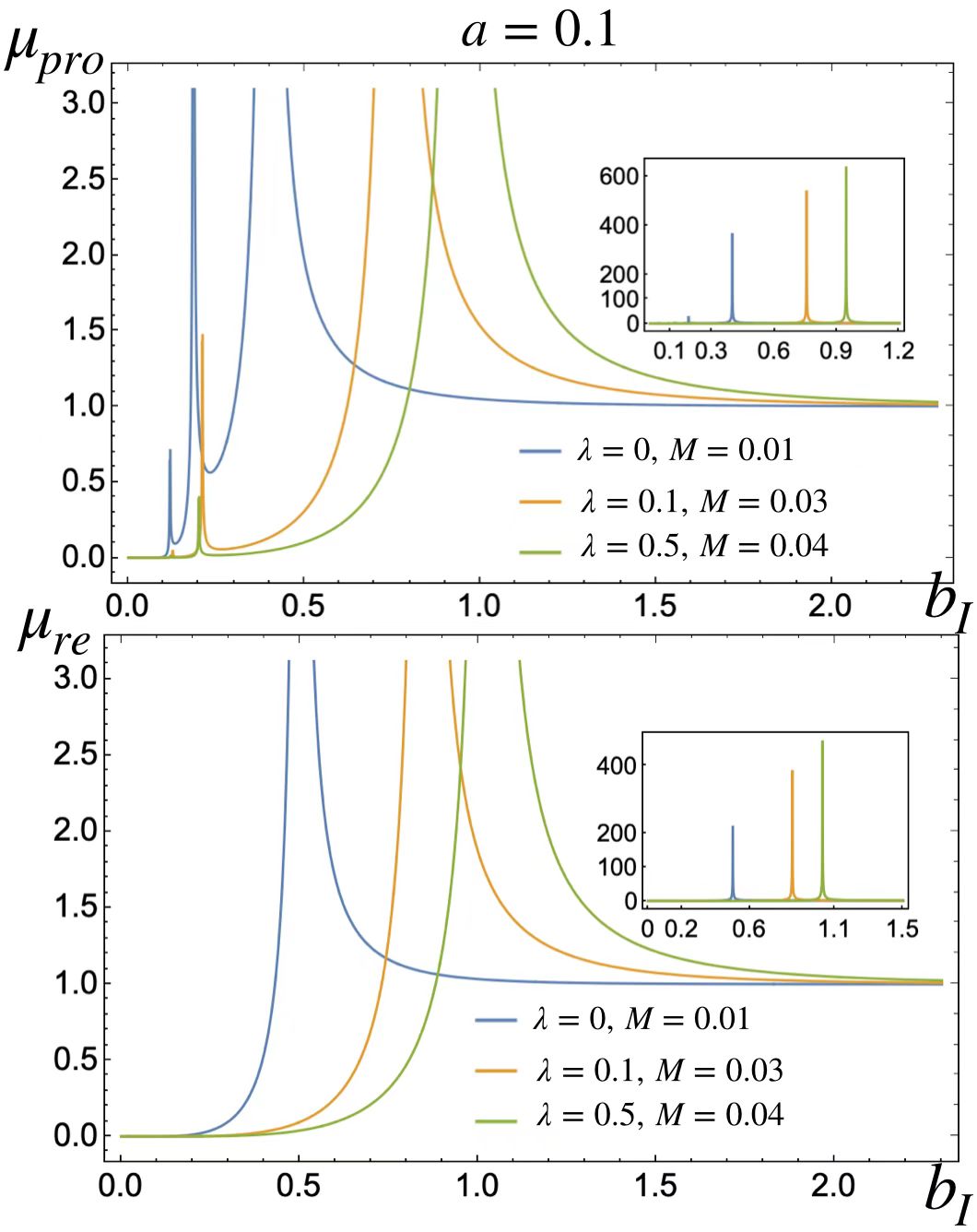}
	\caption{The magnification of Kerr metric  \eqref{eq:kerr metric} with $M \simeq a$. We include the prograde case (upper channel) and retrograde case (lower channel). All of these units have been unified and the value of $0.1~\rm kpc<b_I <2.8~\rm kpc$ ensures the weak field approximation. The maximal order of the magnification is around $10^2$. }
	\label{fig:mag of kerr as a is consistent with m}
\end{figure}

Now let's come to the case $M \simeq a$ where non-trivial results appeared. As we previously discussed, the Kerr BH will become Kerr traversable WH as $a/M>1$, thus the blue line of Fig. \ref{fig:mag of kerr as a is consistent with m} corresponds to the Kerr traversable WH. As we see from Fig. \ref{fig:mag of kerr as a is consistent with m} where we take $M=\mathcal{O}(10^{-2})$, $a=0.1$, the lensing feature for retrograde case is distinctively differeny from the prograde case. While there is only one peak in the retrograde case, there can be at most three peaks in the prograde case. Moreover, the first two gentle peaks for Kerr  traversable WH ($\lambda = 0$, the blue line in figure \ref{fig:mag of kerr as a is consistent with m}) has the magnitude of order $\mathcal{O}(1)$ and $\mathcal{O}$(10), while for the Kerr WH case ($\lambda = 0.1$ and $0.5$, the orange and green line), the corresponding two peaks have the magnitude of order o$(10^{-1})$ and $\mathcal{O}(1)$. However, the magnitude of main peaks for the three cases are all of order $\mathcal{O}(10^2)$. Therefore, we conclude from figure \ref{fig:mag of kerr as a is consistent with m} that, the parameter $\lambda$ suppress the magnitude of gentle peaks greatly while enhance that of main peak gently. 

 Let's address more about the multi-peak feature in the prograde case. A peak is a local maximum in the magnification, i.e.
	\begin{equation}
	\frac{d|\mu|}{db_I} = 0 ~,~ \frac{d^{2}|\mu|}{db_I^2} < 0 ~.
	\end{equation}
	We can estimate the number of peaks in the following way. When one peak exists, there is one stationary point for $|\mu|$. Then, if another peak come into existence, we have two additional stationary points for $|\mu|$, since the additional peak comes with a corresponding valley. Hence, it suffices to count the number of stationary points of $|\mu|$. The condition for stationary point is
	\begin{equation}
	\frac{d|\mu|}{db_I} = 0 ~\to~ \frac{d}{db_I} \left| \frac{\beta}{\theta} \frac{d\beta}{d\theta} \right| = 0 ~.
	\end{equation}
	For simplicity let's take $D_{ls}/D_s \simeq 1$, since it won't change the structure of the equation (of course, in reality $D_{ls}/D_s$ must be smaller than $1$). The lens equation simplifies to $\beta = \theta - \alpha$, and the geometry tells $b_I = D_l \theta \propto \theta$, which means
	\begin{equation}
	\theta = \frac{b_I}{D_l} ~;~ \frac{\beta}{\theta} \frac{d\beta}{d\theta} = \left( 1 - D_l \frac{\alpha}{b_I} \right) \left( 1 - D_l \frac{d\alpha}{db_I} \right) ~.
	\end{equation}
	
	From \eqref{eq:prograde case of kerr} and \eqref{eq:retrograde case of kerr}, the structure of $\alpha(b_I)$ is
	\begin{equation}
	\alpha (b_I) = \frac{b_1}{b_I} + \frac{b_2}{b_I^2} + \frac{b_3}{b_I^3} ~,
	\end{equation}
	where $b_1 > 0$ and $b_3 <0$. For the retrograde case $b_2$ is always positive while for the prograde case $b_2$ can be negative. Again for simplicity let's set $D_l = 1$ and $b_3 = 0$, then 
	\begin{align}
	\label{eq:ddbI}
	\frac{d|\mu|}{db_I} & = \frac{1}{b_I^7} \left( -3b_2b_I^3 + 4b_1^2b_I^2 + 15 b_1b_2 b_I + 12b_2^2 \right) ~.
	\end{align}
	For positive $b_2$, all coefficients in the cubic equations in \eqref{eq:ddbI} are positive except for the $-3b_2b_I^3$ term. Thus, the RHS of \eqref{eq:ddbI} will cross $0$ once. There will be only one stationary point for $\mu$ in $b_2 > 0$ case, and hence only one peak. For negative $b_2$, only the linear term $15b_1b_2 b_I^2$ has a negative coefficient. The RHS can cross $0$ more than once, and hence multiple peaks. 
	
	Surely, to analytically recover the three-peaks, we need to also include the $b_3$ term and expand $d|\mu|/db_I$ with higher orders. However, the expression may be too complicated, and may not be as clear as a direct numerical study. Here we include the simple analytical argument, only to illustrate how multi peaks arise from the behavior of $\alpha$. As we can see, a monotonous $\alpha$ would lead to a monotonous $d\mu/db_I$, and thus only one peak. In the prograde case when $b_2 < 0$, $\alpha$ has more features, which reflects to the expression of $d|\mu|/db_I$, and potentially generates more stationary points of  $d|\mu|/db_I$, and thus more peaks.

 In lensing field, the magnification as an explicit observable for the compact object, its detection will straightfowrdly show the property of the corresponding compact object. According to the current observation for magnification, nearly all of them are showing only one peak of the magnification that leads to the difficulty for distinguishing the objects, such as the wormhole and its associated blackhole. Thus, our theoretical investigation of the multi-peaks for the Kerr-like wormhole (blackhole) will provide a special singal that we could distinguish the them if we can detect the similar curve of magnification in a near future.  To be more precisely, the multi-peaks of magnigication, as showing in Fig. \ref{fig:mag of kerr as a is consistent with m}, tells that the corresponding compact object may be the Kerr-like WH.

\section{RN Case}
\label{sec:RN}

The metric of RN wormhole is found by \cite{Kim:2001ri},
\begin{equation}
	ds^2=-\bigg(1+\frac{Q^2}{r^2}\bigg)dt^2+\bigg(1-\frac{r_0^2}{r^2}+\frac{Q^2}{r^2}\bigg)^{-1}dr^2+r^2d\Omega^2,
	\label{eq:RN metric}
\end{equation}
where $Q$ is the charge and $r_0$ is the radius of the throat.

The RN wormhole \eqref{eq:RN metric} has two specific cases. When $Q=0$, the metric reduces to a massless EBWH with throat radius $r_0$. The $r_0 = 0$ case is more subtle. The metric will have a form 
\begin{equation}
	ds^2 = - \left( 1 + \frac{Q^2}{r^2} \right) dt^2 + \left( 1 + \frac{Q^2}{r^2} \right)^{-1} dr^2 + r^2d\Omega^2,
\end{equation}
similar to a RN black hole with a vanishing mass. However, in this case there's no event horizon since $g_{tt}$ and $g_{rr}^{-1}$ is always regular in the whole spacetime $r>0$. Recall that for a RN black hole
\begin{equation}
	ds^2 = - h dt^2 + h^{-1} dr^2 + r^2d\Omega^2 ~,~ h \equiv 1 - \frac{r_s}{r} + \frac{Q^2}{r^2} ~,
	\label{metric of RN BH}
\end{equation}
the event horizon is defined by the condition $h = 0$, and there's no real solution when $r_s = 0$. Another intuitive way to understand this fact is that, the electric charge term behaves as if exerting a repulsive to a massive object (one may see this by the Newtonian potential $2\Phi = -\frac{r_s}{r} + \frac{Q^2}{r^2}$, and the charge term $r_Q^2$ contributes positively). Therefore if the mass is vanishing, the massive object can only feel repulsive force, thus no event horizon. Since we cannot obtain the metric of RN BH via metric \eqref{eq:RN metric} and we still need to compare the magnification for them. Following the method in Sec. \ref{sec:ellis}, we can firstly obtain the optical Gaussian curvature of RN WH upon to the second order, 
\begin{equation}
	\mathcal{K}_{\rm WH}=\frac{3Q^2-r_0^2}{r^4}-\frac{4r_0^2Q^2}{r^6}+\mathcal{O}(Q^4,r_0^4),
	\label{eq:gaussian cur of RN}
\end{equation}
and its corresponding deflection angle as follows, 
\begin{equation}
	\alpha_{\rm WH}=\frac{r_0^2\pi}{4b_I^2}-\frac{3\pi Q^2}{4b_I^2}+\frac{3r_0^2\pi Q^2}{8b_I^4}.
	\label{eq:deflection angle of RN}
\end{equation}
Once obtained the second order of deflection angle, we can simulate the magnification of RN metric. Then the optical Gaussian curvature of RN WH is derived as follows,
\begin{equation}
	\mathcal{K}_{\rm BH}=\frac{2 Q^4}{r^6}-\frac{3 \left(Q^2 r_0\right)}{r^5}+\frac{3 \left(4 Q^2+r_0^2\right)}{4 r^4}-\frac{r_0}{r^3},
	\label{gaussian cur of RN BH}
\end{equation}
where $r_0=2M$ ($G=1$) in the case of RN BH. Noticing that this optical Gaussian curvature is complete without needing any Taylor expansion, which is different for \eqref{eq:gaussian cur of RN}. Being armed with this Guassian curvature for metric \eqref{metric of RN BH}, then its deflection angle can be straightforwardly obtained as 
\begin{equation}
\alpha_{\rm BH}=-\frac{3 \pi  Q^4}{16 b_{\text{I}}^4}+\frac{4 Q^2 r_0}{3 b_{\text{I}}^3}-\frac{3 \pi  Q^2}{4 b_{\text{I}}^2}-\frac{3 \pi  r_0^2}{16 b_{\text{I}}^2}+\frac{2
	r_0}{b_{\text{I}}}.
\end{equation}

Similar to the Kerr case in section \ref{sec:Kerr}, we study the magnification of RN case with three different parameter setting: $r_0 \gg Q$, $r_0 \sim Q$ and $r_0 \ll Q$.

\begin{figure}[h]
	\centering
	\includegraphics[width=0.93\linewidth]{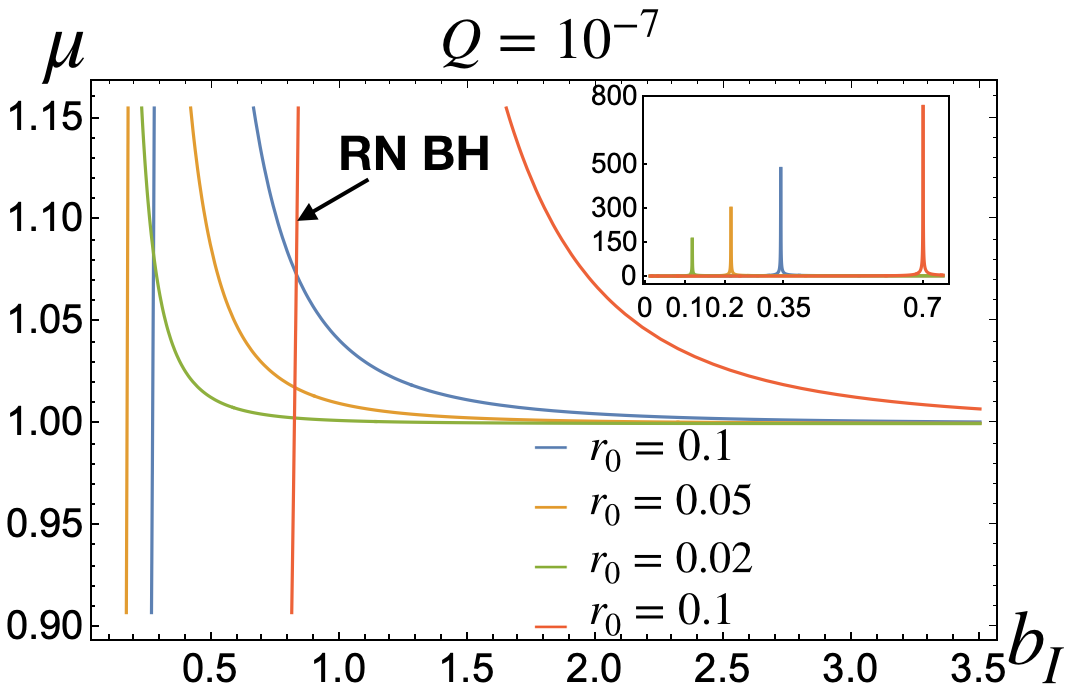}
	\caption{The magnification of metric \eqref{eq:RN metric} for $r_0 \gg Q$. The $b_I$ ranges from $0.1~\rm kpc$ to $2.5~\rm kpc$ keeping the weak field approximation valid. And the red line corresponds to the magnification of RN BH \eqref{metric of RN BH}. } 
	\label{fig:mag of rn as ro is larger than q}
\end{figure}

We start with the $r_0\gg Q$ case, where the metric returns to the massless EBWH when $Q=0$. Figure \ref{fig:mag of rn as ro is larger than q} shows that there's one single peak with $b_I/r_0 = \mathcal{O}(1)$, consistent with the result from massless EBWH. The red line corresponds to the RN BH ($r_0$ denotes the mass), our numerical simulations reveals that the order of magnification of RN BH will be larger than the case of RN WH as $r_0$ is compatiable. 

\begin{figure}[h]
   	\centering
   	\includegraphics[width=0.93\linewidth]{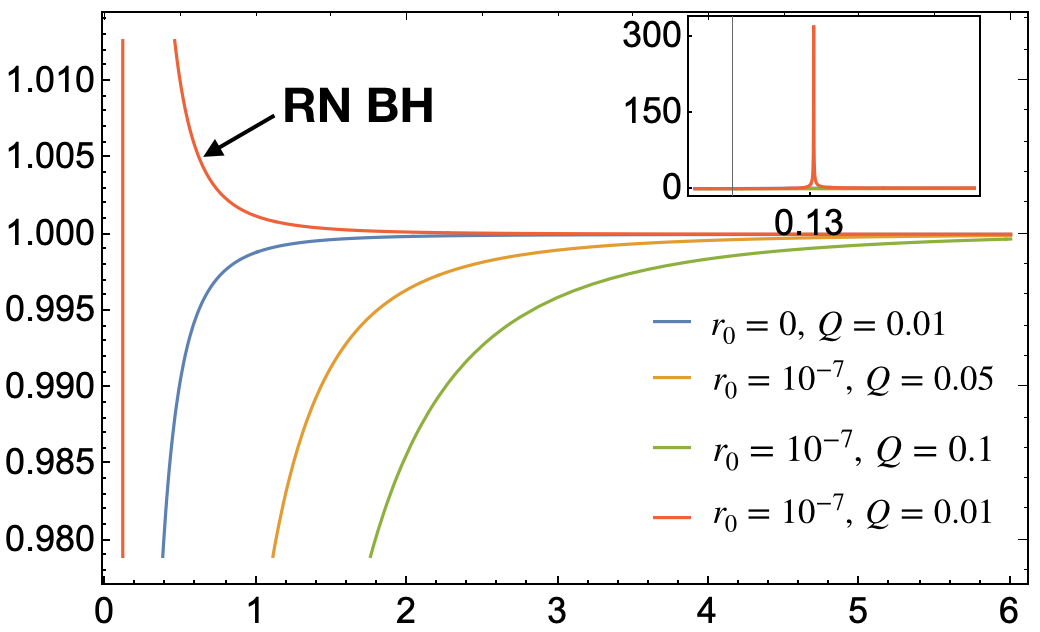}
   	\caption{The magnification of metric \eqref{eq:RN metric} for $r_0 \ll Q$. The $b_I$ ranges from $0.1~\rm kpc$ to $6~\rm kpc$ keeping the weak field approximation valid. The blue line with $r_0 = 0$ corresponds to a RN black hole with vanishing mass. And the red line corresponds to the magnification of RN BH \eqref{metric of RN BH}.}
   	\label{fig:mag of rn as q is larger than r0}
\end{figure}

\begin{figure}[h]
   	\centering
   	\includegraphics[width=0.93\linewidth]{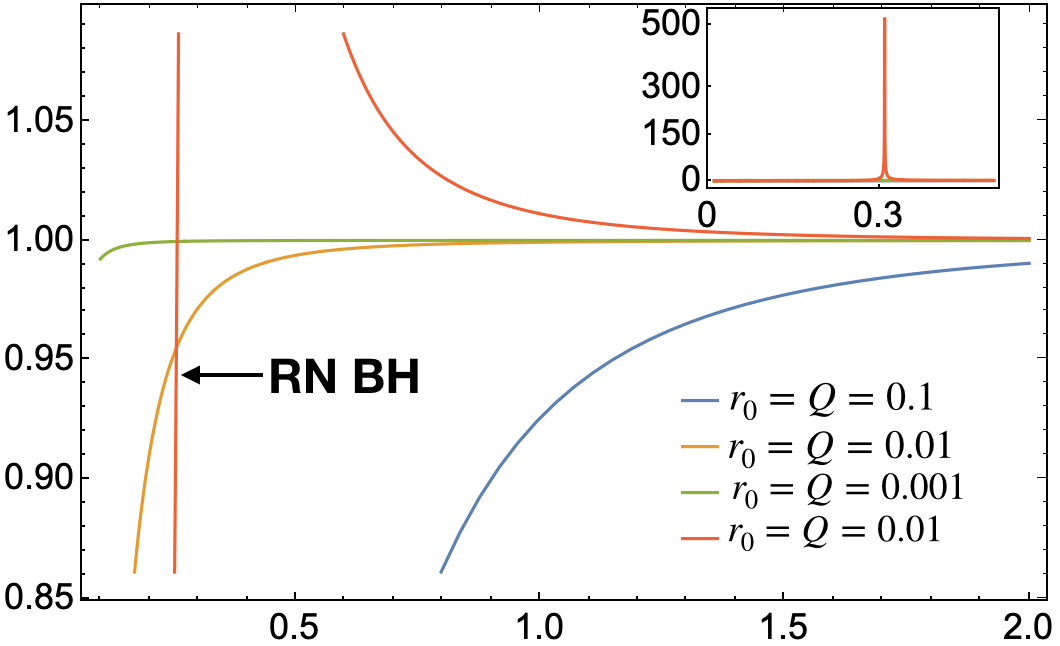}
   	\caption{The magnification of metric \eqref{eq:RN metric} for $r_0 \simeq Q$. The $b_I$ ranges from $0.1~\rm kpc$ to $2~\rm kpc$ keeping the weak field approximation valid. And the red line corresponds to the magnification of RN BH \eqref{metric of RN BH}.}
   	\label{fig:mag of rn as q is comparable with r0}
\end{figure}

The rest two cases show similar results. We firstly give the analysis of RN WH. We plot the magnification for $r_0 \ll Q$ in figure \ref{fig:mag of rn as q is larger than r0}, and $r_0 \simeq Q$ in figure \ref{fig:mag of rn as q is comparable with r0}, and both figures show the same trend. There's almost no magnification effect for the case of RN WH. Besides, the images will be demagnified for smaller $b_I$. However, when $b_I$ approaches $r_0$, we might need to worry about higher-than-second order contributions like $r_0^3/b_I^3$, so we shall treat the small $b_I$ part less trustworthy. The magnification will be different of RN BH comparing with RN WH. From Figs. \ref{fig:mag of rn as q is larger than r0} and \ref{fig:mag of rn as q is comparable with r0}, one can clearly see that there is a peak denoting there is obivous magnified image in some certain scales. If one can determine the compact objects as the RN WH or RN BH, one can distinguish them via the observations. 

In conclusion, we find the RN metric with $r_0 \gg Q$ shows similar lensing effect as a massless EBWH, while for $Q \geq r_0$ there's almost no magnification effect for RN WH. However, RN BH will appear the peak of their corresponding magnification. In some sense, one can tell apart the RN WH and RN BH. Furtherly, it is still hard to distinguish RN BH from the other generic BHs.

\section{Conclusion}
\label{sec:conclusion}
We explore the microlensing effects of a wormhole with a relatively strong presence of gravity. Although we still work in the weak field approximation, we investigate the deflection angle in the second order. The three typical wormholes (blackholes), the Schwarzschild WH/BH, Kerr WH/BH, and RN WH/BH, are investigated. 

 We find that it is possible to distinguish the Kerr wormhole from other compact objects. More specifically, in the prograde case, there will be multi-peaks when the contributions from mass and angular momentum are comparable. As shown in figure \ref{fig:mag of kerr as a is consistent with m}, the ratios between the magnitude of gentle peaks and main peak for Kerr WH ($\lambda$ is not zero) and Kerr traversable WH ($\lambda=0$) are different. One can also distinguish the RN WH and RN BH via their corresponding magnification since there will be one peak for RN BH, the case of RN WH is negative as $r_0\ll Q$ or $r_0=Q$. The other cases are hard to distinguish, since the behavior of magnifications of wormholes and corresponding blackholes are similar, and the only difference is the magnitude of magnification and the location of peaks. 

Our work is a preliminary check on this topic. There are many interesting ideas to be studied in the future. Firstly, to fully address the issue, we need to go to the strong field limit. Then, in the strong field case, it is possible that the non-trivial topology influence not only the deflection angle, but also the lens equations. Hence, we need to improve the techniques in this paper. Secondly, we only studied selected models, and we need more examples to strengthen our conclusion. For example, the no-hair theorem \cite{Bekenstein:1971hc,Bekenstein:1995un} tells that blackholes are uniquely determined by their mass, charge, and angular momentum. It is interesting to study wormholes and blackholes sharing these three same parameters. Finally, GBT may fail for certain modified gravity theories, such as the gravitational theory with a modification to the Gauss-Bonnet term \cite{Glavan:2019inb}, can we improve the GBT formalism in this situation?

 \section*{Acknowledgements}
 We appreciate the stimulating discussions with Bichu Li, Yuhang Zhu.  
 LH and KG are funded by NSFC grant NO. 12165009. MZ is funded by grant NO. UMO-2018/30/Q/ST9/00015 from the National Science Center, Poland.





\end{document}